\documentclass[10pt,letterpaper]{article}
\pdfoutput=1 

\usepackage{cite}
\usepackage{amsmath}
\usepackage{amsfonts}
\usepackage{amssymb}
\usepackage{graphicx}
\usepackage[left=25mm,right=25mm,top=25mm,bottom=25mm]{geometry}
\usepackage{float}
\usepackage{subfigure}
\usepackage{authblk}
\usepackage[margin=3eM]{caption}
\usepackage[colorlinks=true, linkcolor = blue, urlcolor  = blue, citecolor = blue, pdfborder={0 0 0}, unicode = true, linktoc=all]{hyperref}%
\begin{document}

\title{Saturation in regular, exotic and random pore networks}

\author{M\'at\'e Benj\'amin Vizi}
\author{P\'eter \'Arp\'ad Mizs\'ak}
\author{Tam\'{a}s Kalm\'{a}r-Nagy$^\dagger$}
\affil{$^\dagger$Department of Fluid Mechanics, Faculty of Mechanical Engineering, Budapest University of Technology and Economics, Budapest, 1111, Hungary; kalmarnagy@ara.bme.hu}

\renewcommand\Authands{ and }

\date{}
\maketitle

\maketitle

\begin{abstract}
Porcolation simulations were carried out on various networks; both regular and irregular.
The saturation curve was obtained for cubic networks, localized and completely random 3D networks and networks based on exotic graphs like Sierpi\'nski triangle and carpet. 
For the random graph generation a modification of the cell list algorithm was introduced, which is capable of generating local random graphs efficiently. 
With the help of this graph generation method, the effect of locality was investigated, and it was proven to be an important property of random networks from the viewpoint of liquid propagation. 
The saturation curves of local random networks with different prescribed pore degree distributions were also obtained.
\end{abstract}

\section{Introduction}

Liquid propagation in porous medium is a much-studied topic. Research in this
area is important, for example, for the petroleum industry
\cite{chavent1986mathematical,coats1998compositional}, for groundwater
contamination studies \cite{bear2012modeling,pinder1973galerkin} and for road
construction \cite{roseen2011water}.

A porous medium can be imagined to have pores (cavities) and capillaries
(throats) connecting the pores. The invading fluid needs different external
pressure to enter into different sized throats or pores. This so-called entry
pressure can be calculated from the Washburn equation
\cite{washburn1921dynamics}%

\begin{equation}
p=-\dfrac{2\gamma\cos(\theta)}{\rho}, \label{eq_washburn2}%
\end{equation}
where $\gamma$ is the surface tension of invading phase, $\theta$ is the
contact angle between the non-wetting invading phase and the material and
$\rho$ is the characteristic radius of the capillary (or throat).

There are two main computational approaches to model liquid propagation in
porous materials \cite{sahimi2011flow}: the continuum and the pore network
approach. Commercial packages like Fluent and Comsol utilize the continuum
approach, in which the porous material is treated as a volume-averaged
continuum. The fairly low computational cost of this approach however means
that the microscale features of the material are not resolved, limiting the
method to problems in which the connectivity of the pore space does not play a
major role.

Pore network (discrete) modeling resolves the microscale features of the
medium at the expense of larger computational cost. The porous medium can be
modeled as a graph, where the vertices and edges correspond to the pores and
capillaries, respectively. These pore-scale models date back to the work of
Fatt \cite{fatt1956network1,fatt1956network2,fatt1956network3}. The transport
inside the network is modeled using finite difference schemes. This approach
is widely used to simulate the multiphase flows in fuel cell electrodes
\cite{putz2013openpnm}. OpenPNM (an
open-source pore network modeling package) \cite{OpenPNM}
also applies this approach. The advantages of pore network modeling compared
to continuum approach is presented in \cite{openpnm2013,openpnm2016}.

The distribution of pore sizes is a crucially important property of porous
materials. Mercury porosimetry \cite{giesche2006mercury} is a commonly used
method to determine the pore size distribution of rock samples. During this
process mercury is forced into the samples using increasing external pressure.
\ The volume of the injected mercury as the function of the pressure is the
so-called saturation curve.

The modeling of mercury porosimetry was first studied by Chatzis and
Androutsopoulos
\cite{chatzis1977modelling,androutsopoulos1979evaluation,chatzis1985modeling}.
An external pressure driven access-limited invasion percolation model called porcolation was introduced in \cite{bak2016porcolation}.

The porcolation method can be used for real networks provided that the statistical information required for the network generation is available.
According to recent studies \cite{tahmasebi2017image}, the modeling of granular porous media can be effectively and accurately done based on processing \mbox{2D/3D} images.

The main objective of the current work is to study saturation properties of both regular and irregular networks. 
We carried out porcolation simulations on square/cubic networks, networks based on exotic graphs like the Sierpi\'{n}ski triangle and carpet, and also on localized and completely random networks. 
Saturation curves were determined with OpenPNM. 

This paper is structured as follows:
in Section \ref{section:simulation_algs} the theoretical background of different
percolation models is presented.  Section \ref{section:porcolation} presents saturation curves of porcolation simulations on square and cubic networks. 
In Section \ref{section:exotic-network} saturation curves for Sierpi\'{n}ski triangle and Sierpi\'{n}ski carpet networks are shown. 
In Section \ref{section:irregular-graphs} the locality properties of irregular 3D networks were investigated with a newly developed graph generation model and saturation curves were also obtained for networks with different pore degree distribution. 
Section \ref{section:conclusion} concludes the paper.


\section{Percolation models}\label{section:simulation_algs}

Percolation theory was introduced in 1957 by Broadbent and Hammersley
\cite{BroadbentHammersley1957Percolation}. They investigated how the random
properties of a medium influence the percolation of a fluid through it. In the
following four percolation models are presented.

\subsection{Ordinary Percolation}

\begin{figure}[th]
	\centering
	\subfigure[]{
		\includegraphics[width=0.25\textwidth]{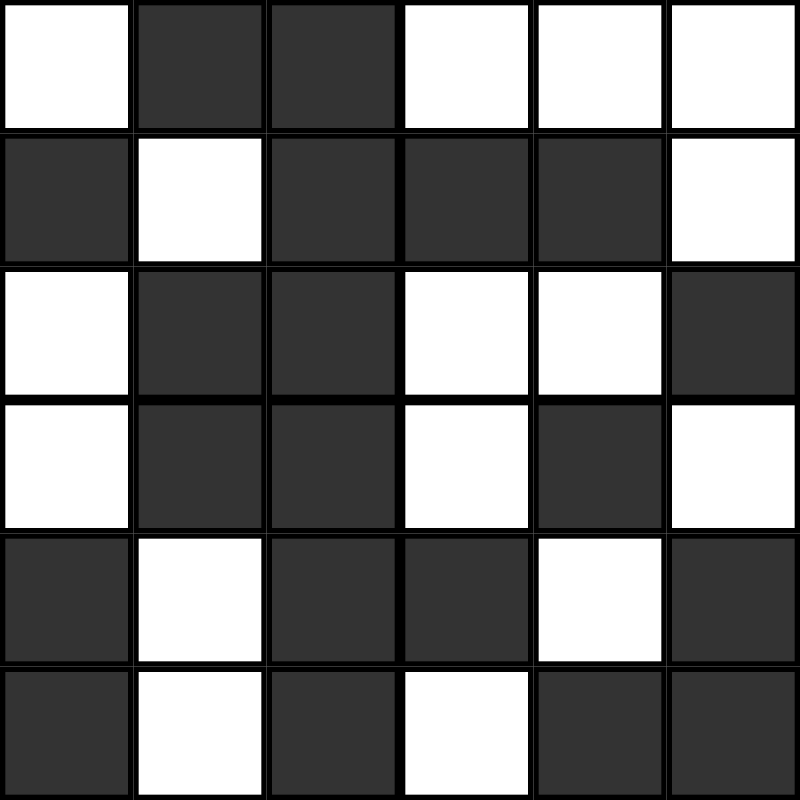}
		\label{fig:lattice-site}
	}
	\quad
	\subfigure[]{
		\includegraphics[width=0.25\textwidth]{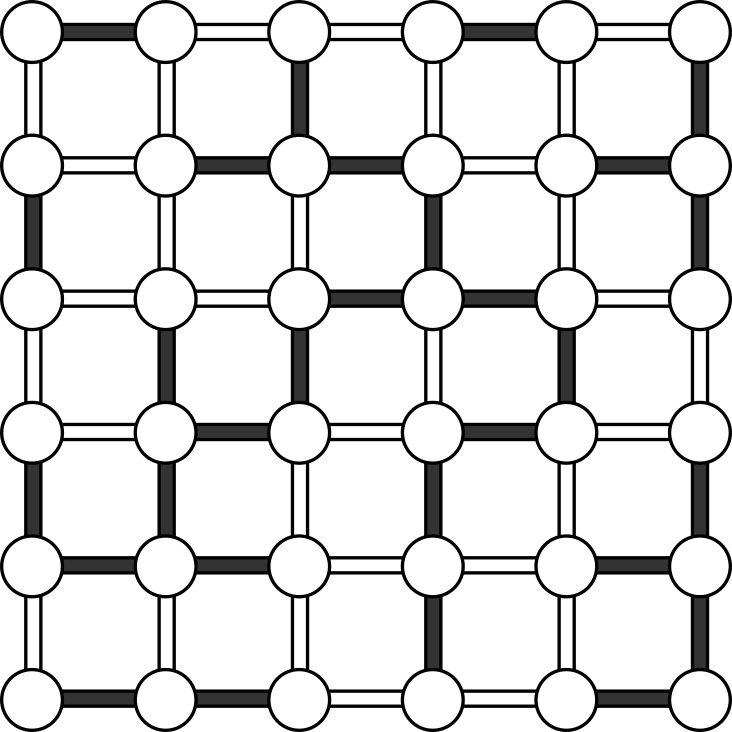}
		\label{fig:lattice-bond}
	}
	\caption{(a) Site-percolation, (b)  Bond-percolation on square lattice.}%
	\label{fig:lattice}%
\end{figure}

There are two fundamentally different types of the ordinary percolation model:
bond-percolation and site-percolation \cite{christensen2002percolation}
(Figure \ref{fig:lattice}).

In \textbf{site-percolation} (Figure \ref{fig:lattice-site}) each lattice site
is occupied with some probability $P$.
Occupied sites having one common side are called neighbors \cite{percolation1994book}, while a group of neighboring occupied sites is called a cluster. Clusters have a crucial role in percolation theory, since the existence of spanning cluster (a cluster that
connects opposite boundaries) means that the invading fluid (the fluid that
enters the medium under pressure) can percolate through the medium. 
If the occupation probability $P$ is small, there is only a slight chance of having a
spanning cluster. On the other hand, if $P$ is nearly 1, there will almost
certainly be a spanning cluster. 
The critical value of occupation probability
($P_\text{crit}$), at which an infinite cluster appears in an infinite lattice is $0.593$ \cite{gebele1984site} and $0.312$\cite{grassberger1992numerical} for two dimensional square and three dimensional cubic networks, respectively.

In \textbf{bond-percolation} (Figure \ref{fig:lattice-bond}) it is not the
lattice sites, but the connecting bonds that are occupied with probability $P$.
Site- and bond-percolation yields different critical probabilities for the same lattice, but the values can be calculated from each other \cite{Berg1982Note,fisher1961some}.

\subsection{Invasion Percolation}

The existence of a spanning cluster in ordinary percolation is a static
property, thus ordinary percolation does not say anything about the dynamics of
cluster growth (i.e. liquid propagation). Invasion percolation was introduced
in \cite{lenormand1980description,wilkinson1983invasion}, as a variant of the
ordinary percolation to fulfill the need of describing the dynamics of liquid
propagation in porous medium. Invasion percolation can also be site- or bond-based.

The basic idea of invasion percolation is that every site (or bond) has an
invasion resistance value $r \in{} [0,\,1]$. The invading phase starts from a
prescribed region (set of sites), and at every step it occupies the most
easily \textquotedblleft accessible\textquotedblright\ site, i.e. the site
that is a neighbor of an already invaded site with the lowest resistance.

\subsection{Porcolation and Drainage}\label{subsection:por_drain}

The porcolation model (PORisometry perCOLATION) is an access limited
site-percolation model introduced in \cite{bak2016porcolation}. The idea of
this model came from porosimetry experiments, where the injection pressure
of the invading non-wetting fluid is gradually increased. The sites from where the fluid is injected into the medium are called the starting set.

\begin{figure}[H]
	\centering
	\includegraphics[width=0.9\textwidth]{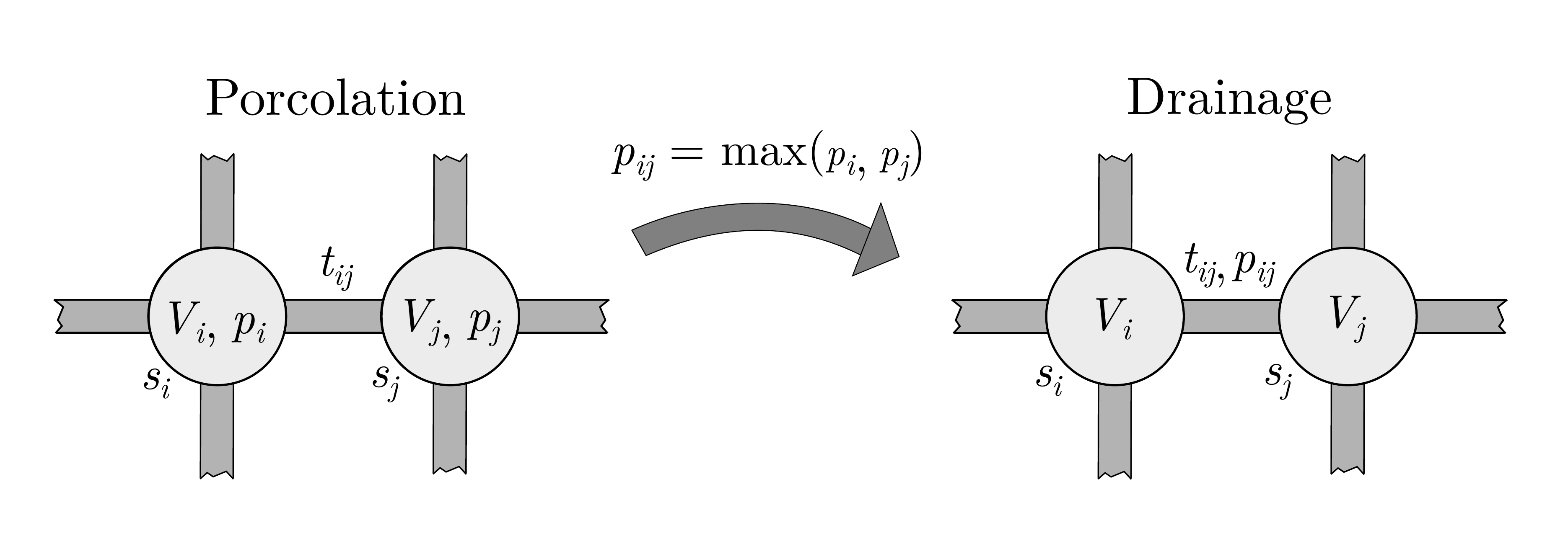}
	\caption{Mapping a porcolation graph to a drainage graph.  $t_{ij}$ is the throat connecting vertices
		$s_{i}$ and $s_{j}$. $V_{i}$ and $V_{j}$ are the pore volumes, while $p_{i}$
		and $p_{j}$ are the corresponding pore entry pressure values. The calculated
		throat entry pressure is $p_{ij}=\max(p_{i},\,p_{j})$
	}%
	\label{fig:porcolation-drainage}%
\end{figure}

In the porcolation model, each site $s_i$ has a volume $V_{i}$ and an invasion resistance (entry pressure value $p_{i}$) as shown in Figure \ref{fig:porcolation-drainage}. 
The $t_{ij}$ throat defines the connection between $s_i$ and $s_j$ sites.
This process is driven by an external pressure
$p\in{}[0,\,1]$. 
For a given pressure $p$, all sites with invasion resistance
$p_i\leq{}p$ get occupied, provided they are connected to the starting set
through a chain of neighboring sites having resistances $p_i\leq{}p$. Hence,
the main difference from invasion percolation is that in porcolation all
accessible sites can be invaded simultaneously.

Drainage is an access limited bond-percolation model. In drainage, sites $s_i$ are connected by throats $t_{ij}$, which have individual entry pressure values $p_{ij}$. In this case the accessible throats are the ones that are connected
with already occupied bonds to the starting set. A site
that has a connecting occupied bond is also occupied instantly.

The mapping from a porcolation graph to a drainage graph (also shown in Figure~ \ref{fig:porcolation-drainage}) is quite simple by assigning to a bond the maximum of the two connected pore entry pressure values, i.e.
\begin{equation}
p_{ij}=\max(p_{i},\,p_{j}). \label{eq_porcolarion-drainage}%
\end{equation}

\noindent
The physical meaning of this equation is that a throat becomes occupied only if both connected pores become occupied.

\section{Porcolation simulations with OpenPNM \label{section:porcolation}}

The total volume of pores in the porcolation model is (the index $i$ runs through all the pores)
\begin{equation}
V_\mathrm{total} = \sum_{i} V_i.
\end{equation}

\noindent The saturation is the ratio of occupied volume and total volume

\begin{equation}
S(p) = \frac{1}{V_\mathrm{total} } \sum_{j} \ V_{j}, \quad \text{for all $j$ with $p_j \leq p$ and $v_j$ is accessible from the starting set.}
\end{equation}

The porcolation simulations were carried out in OpenPNM  whose built-in drainage model is used with the correspondence described in Section \ref{subsection:por_drain}, i.e. the throat entry pressure values $p_{ij}$ were obtained by Equation \eqref{eq_porcolarion-drainage}.

\subsection{Validation on regular graphs}

\begin{figure}[th]
	\centering
	\subfigure[]{
		\includegraphics[width=0.25\textwidth]{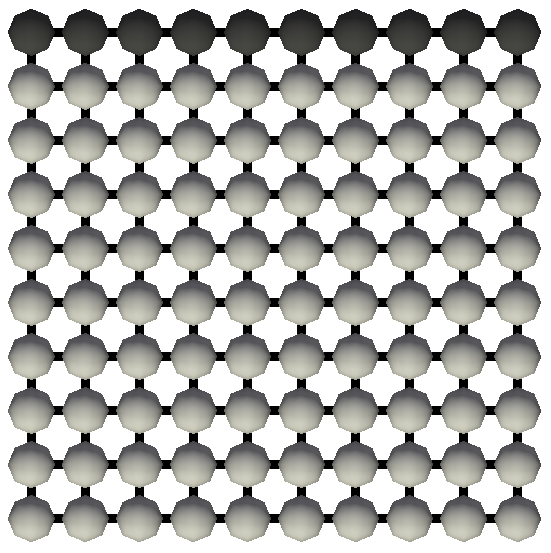}
		\label{fig:one_side_percolation}
	}
	\quad
	\subfigure[]{
		\includegraphics[width=0.25\textwidth]{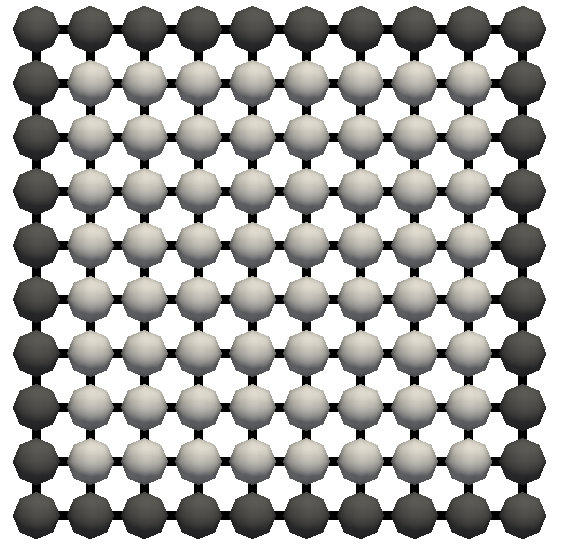}
		\label{fig:four_side_percolation}
	}
	\caption{Porcolation networks with different starting sets: top side of the lattice (a) and full boundary of the lattice (b).}%
	\label{fig:diff_side_percolation}%
\end{figure}

Porcolation simulations were carried out on  $1000 \times1000$ square and $100 \times100 \times100$ cubic lattices ($10^6$ vertices for both). 
The pore entry pressure
values $p_i$ were independently, uniformly generated from $[0,\,1]$. Unit volume
was assigned for each pore, i.e. $V_i = 1$.
Two different starting sets were considered for both the two dimensional and three dimensional cases (Figure
\ref{fig:diff_side_percolation}). 

Fifty equidistant pressure steps were taken in the $p \in [0,\,1]$ range. 
100 simulations were run for each case taking about 3 hours on a 3\textsuperscript{rd} generation, $3.2$ GHz Intel processor; the CPU time depends only on the number of pores and connections, it is independent of the dimension of the graphs. Since the pores have unit volume, the saturation for a given pressure $p$ is simply the ratio of the number of
occupied pores and all pores. 

\begin{figure}[b!]
	\centering
	\includegraphics[width=0.7\textwidth]{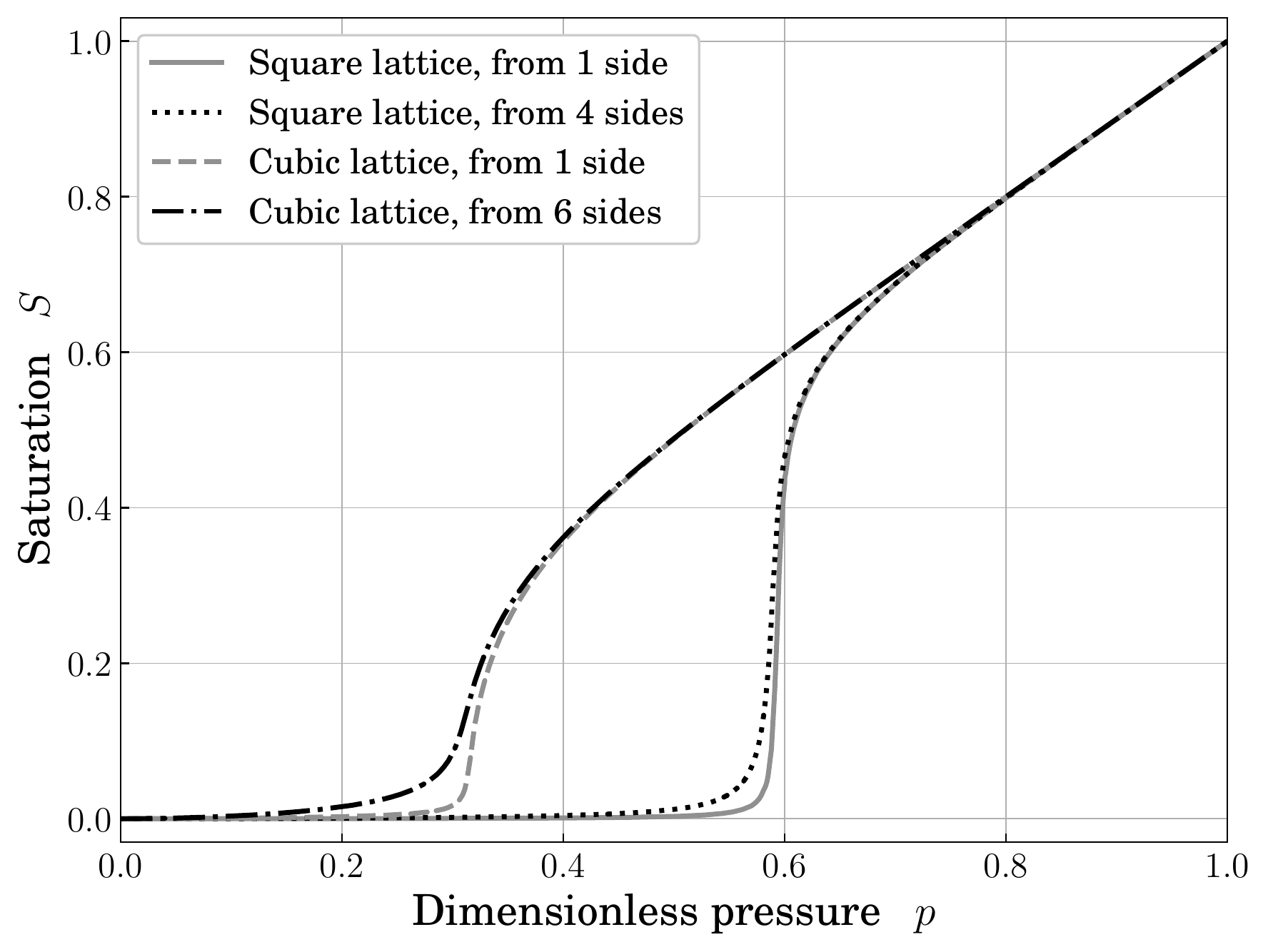}
	\caption{Saturation curves for square and cubic networks}%
	\label{fig:porcolation_large}%
\end{figure}
\begin{figure}[b!]
	\centering
	\includegraphics[width=0.7\textwidth]{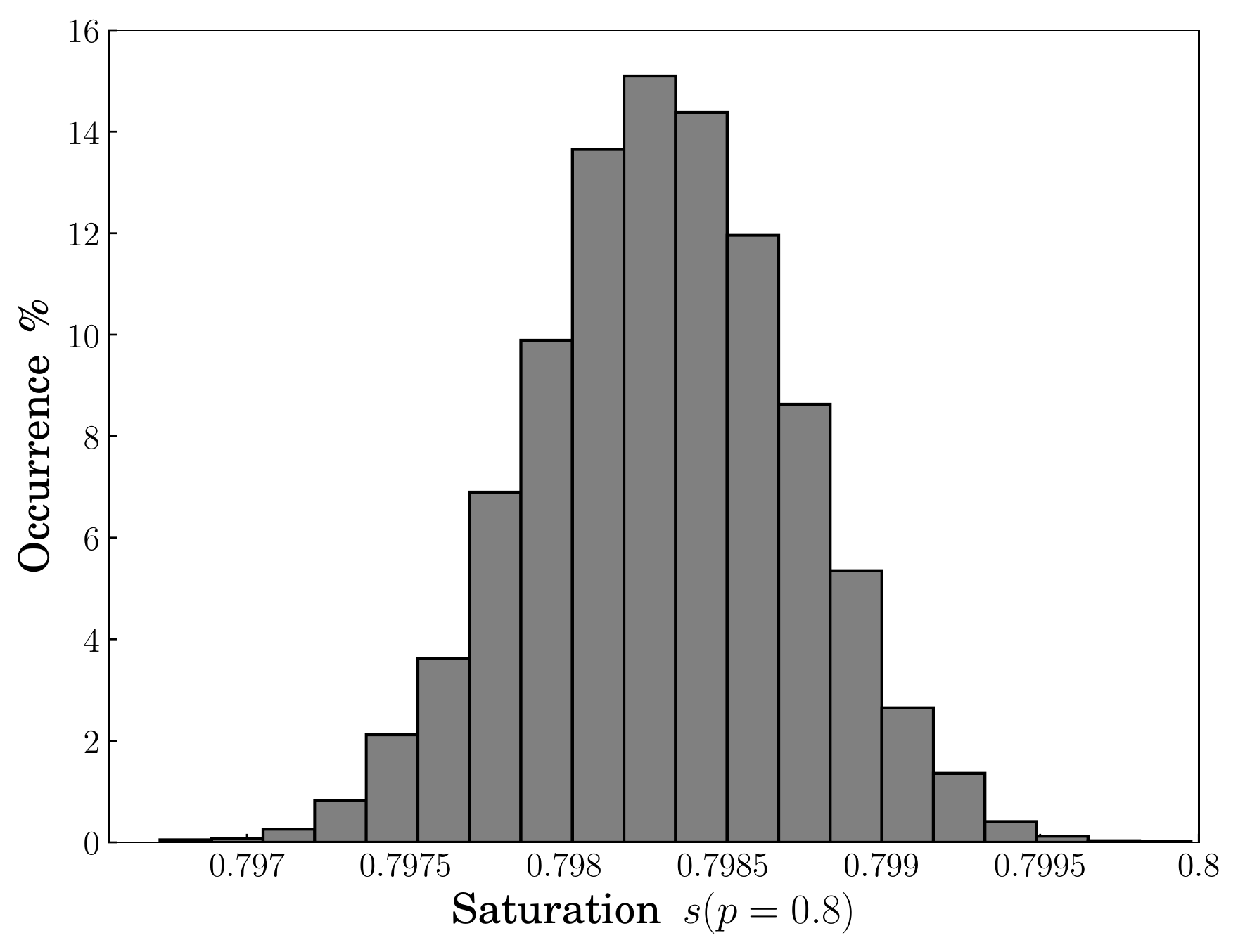}
	\caption{The histogram of saturation for square network with one-sided invasion at $p=0.8$}%
	\label{fig:porcolation_fluc_5}%
\end{figure}

The average saturation curves are shown on
Figure \ref{fig:porcolation_large}. The inflection points of these saturation
curves correspond to the critical probability ($P_\text{crit}$) of ordinary
percolation.
The inflection point for the $1000^2$ square network for one-sided porcolation is $0.592$ (0.16\% difference from the theoretical 0.593). For the $100^3$ cubic network the inflection point is $0.308$ (1.3\% difference from the theoretical 0.312).

\noindent The inflection point of the $1000^2$ square network in the
case of four-sided invasion is around $0.596$~($0.51$\%  difference) and the inflection point of the
$100^3$ cubic network in the case of six-sided invasion is
around $0.309$~($0.96$\%  difference).
Figure \ref{fig:porcolation_fluc_5} shows the histogram of the saturation values for $p=0.8$ (square lattice, one-sided).

\clearpage
\section{Porcolation on exotic graphs}\label{section:exotic-network}

\begin{figure}[h]
	\centering
	\subfigure[]{
		\includegraphics[height=3cm]{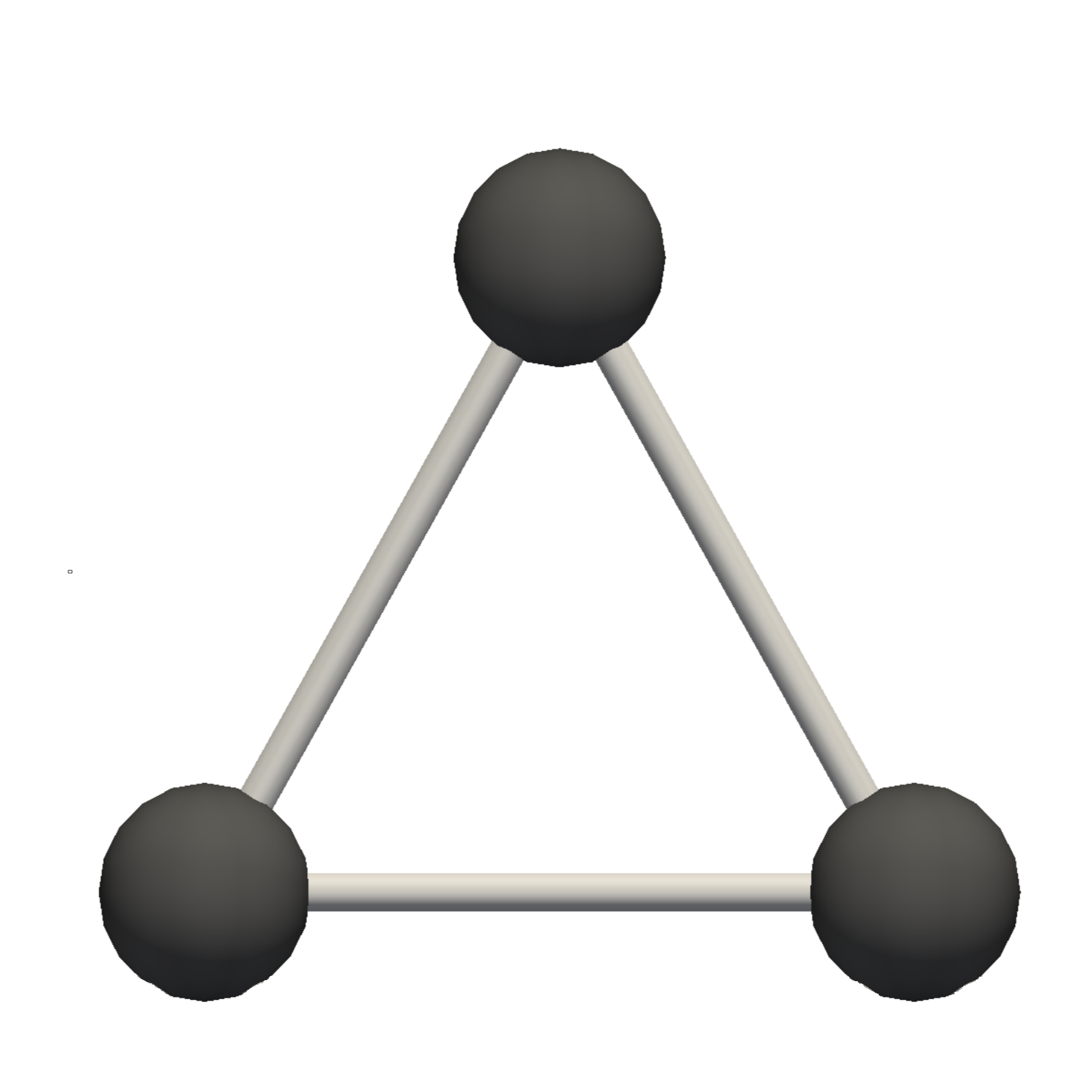}
		\label{fig:sierpenski-triangle-2}
	}
	~
	\subfigure[]{
		\includegraphics[height=3cm]{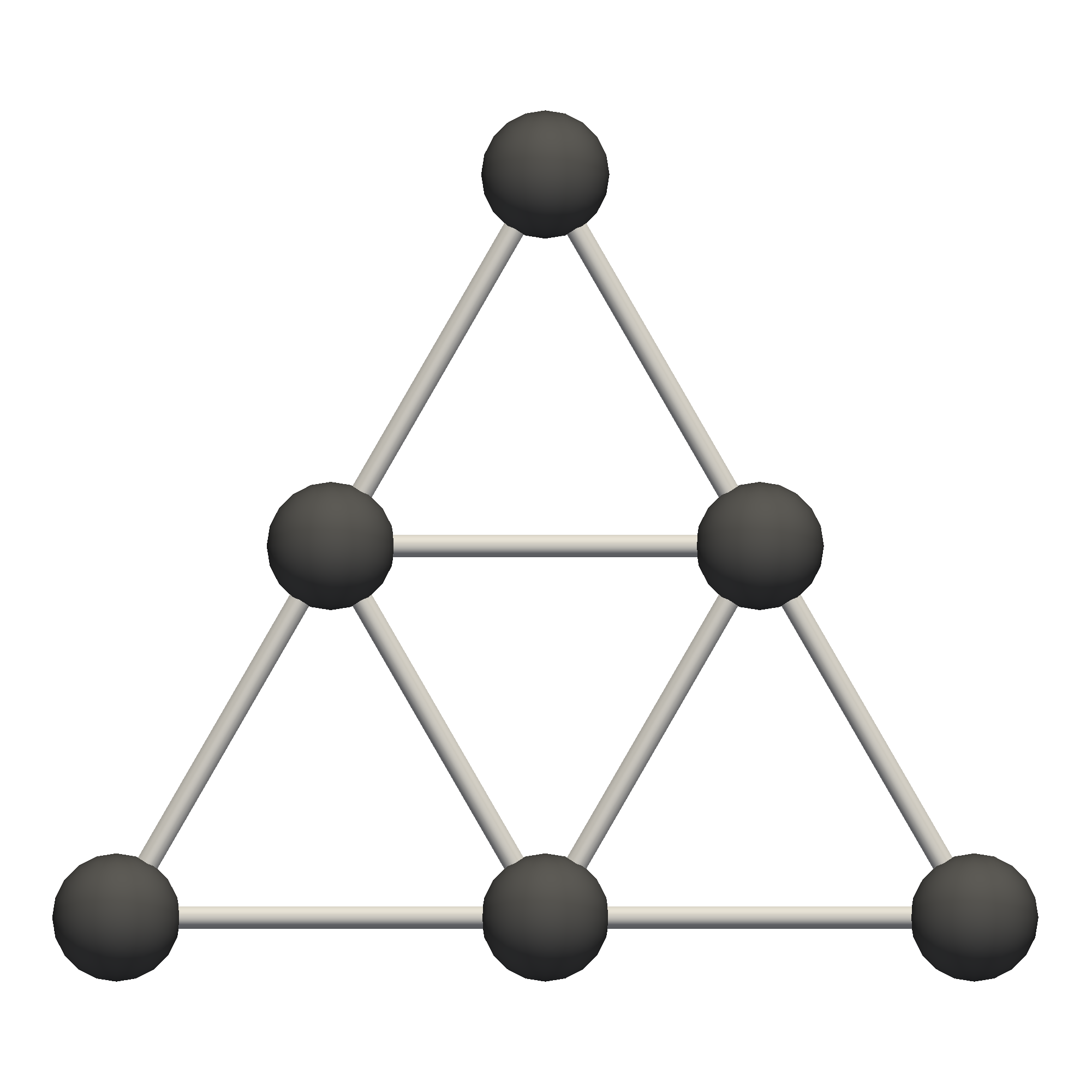}
		\label{fig:sierpenski-triangle-3}
	}
	~
	\subfigure[]{
		\includegraphics[height=3cm]{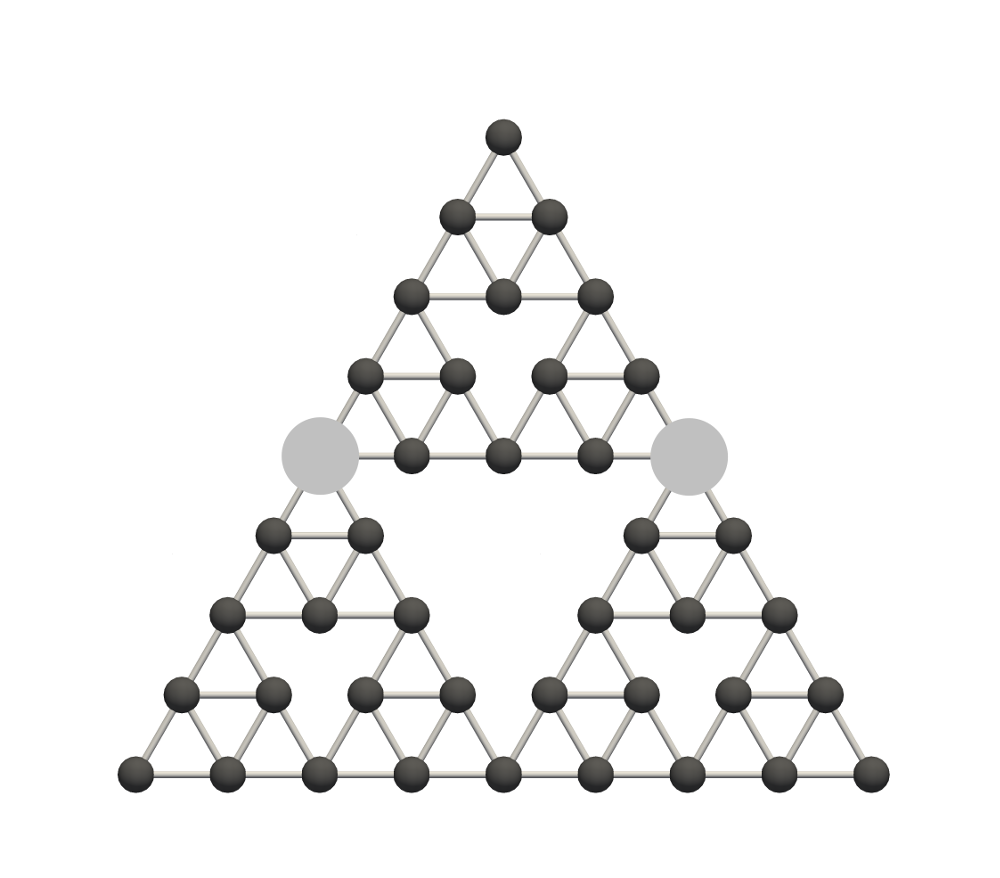}
		\label{fig:sierpenski-triangle-4}
	}
	\caption{The networks based on Sierpi\'nski triangle for different graph levels: (a) $0^\text{th}$ level, (b) $1^\text{st}$ level, (c) $3^\text{rd}$ level }%
	\label{fig:sierpenski-triangle-network}%
\end{figure}

\begin{figure}[h]
	\centering
	\subfigure[]{
		\includegraphics[height=3cm]{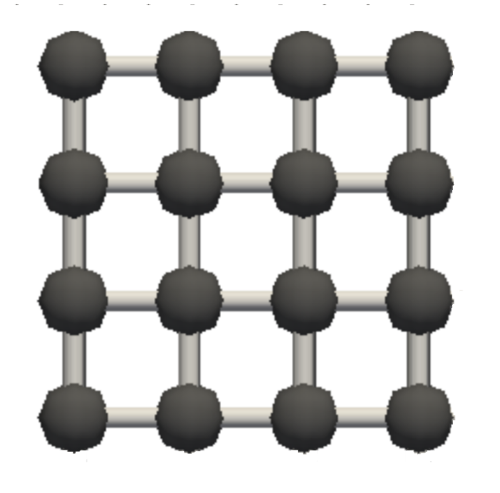}
		\label{fig:sierpenski-carpet-2}
	}
	~
	\subfigure[]{
		\includegraphics[height=3cm]{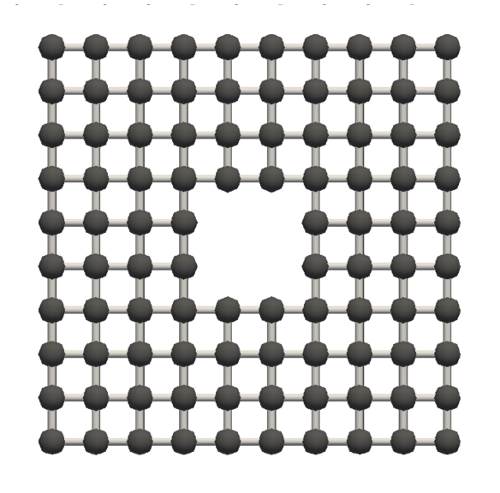}
		\label{fig:sierpenski-carpet-3}
	}
	~
	\subfigure[]{
		\includegraphics[height=3cm]{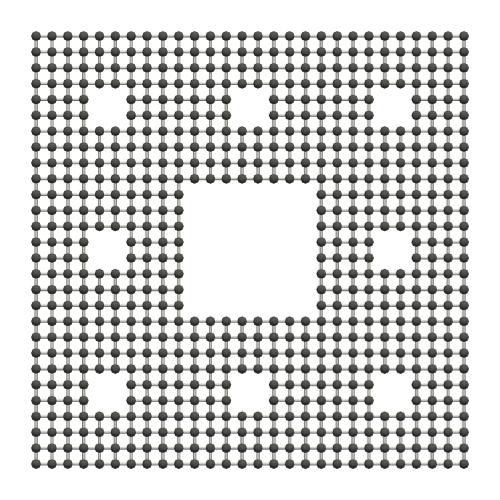}
		\label{fig:sierpenski-carpet-4}
	}
	\caption{The networks based on Sierpi\'nski carpet for different graph levels: (a) $1^\text{st}$ level, (b) $2^\text{nd}$ level, (c) $3^\text{rd}$ level }
	\label{fig:sierpenski-carpet-network}%
\end{figure}

We investigated porcolation on Sierpi\'{n}ski triangle and Sierpi\'{n}ski
carpet style graphs. 
Percolation simulation on the Sierpi\'{n}ski carpet were applied for financial calculations in \cite{pei2015volatility}. 
Finite realizations of Sierpi\'{n}ski triangle and Sierpi\'{n}ski carpet style graphs are shown in Figures \ref{fig:sierpenski-triangle-network} and \ref{fig:sierpenski-carpet-network}, where the level of the graph is the number of iterations required to build the graph from entities of the previous level.

For the simulations the entry pressure values were generated from a uniform distribution in $[0,\,1]$ and the pore volumes were taken as unity. The starting set was the bottom side of the graphs.

\begin{figure}[h]
	\centering
	\subfigure[]{
		\includegraphics[width=0.6\textwidth]{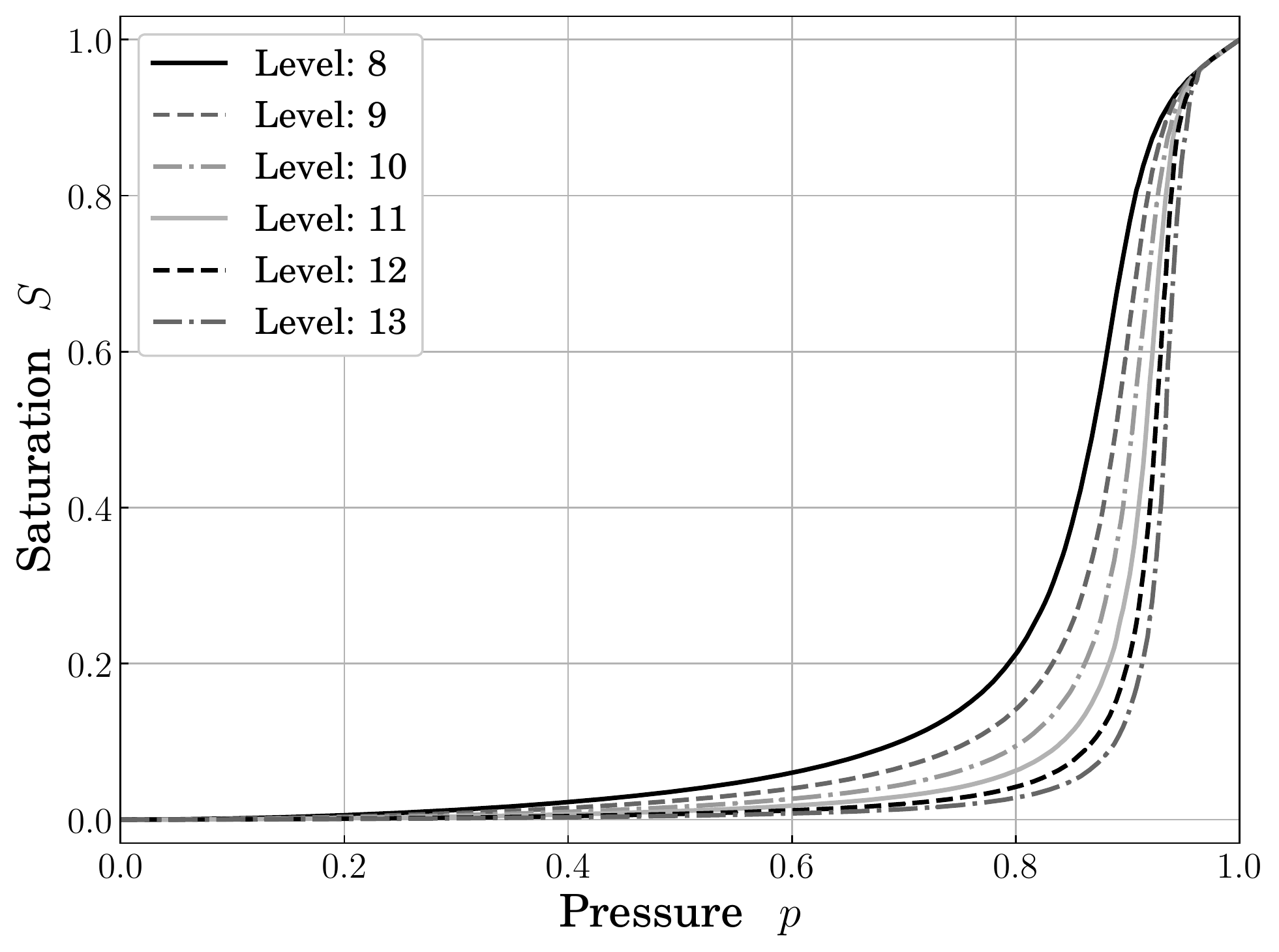}
		\label{fig:sierpenski-triangle-sat}%
	}
	~
	\subfigure[]{
		\includegraphics[width=0.3\textwidth]{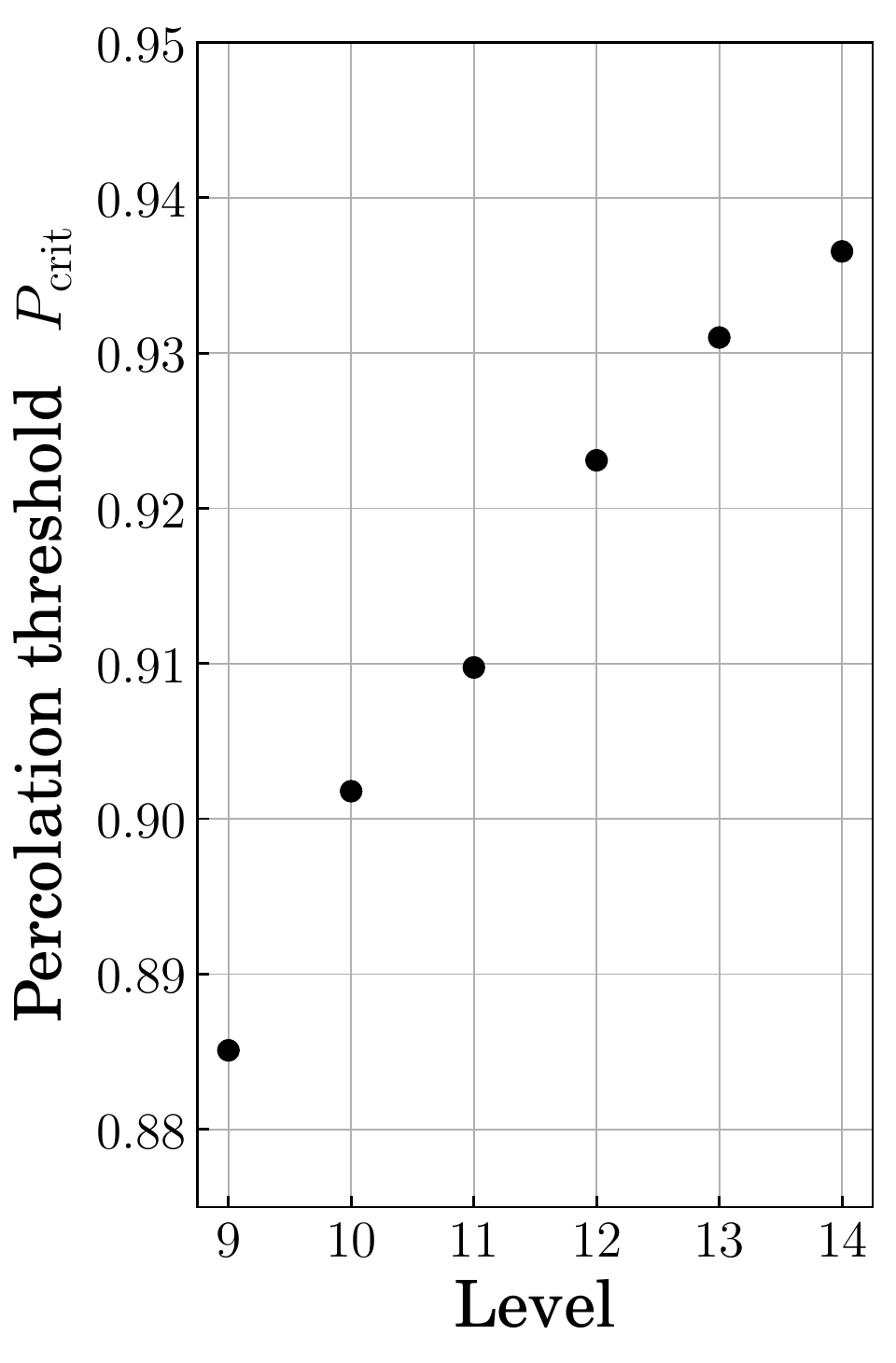}
		\label{fig:sierpenski-triangle-inf}
	}
	\caption{(a) Saturation curves for Sierpi\'nski triangles of different levels,  (b) Percolation thresholds vs. graph level }%
	\label{fig:sierpenski-triangle-results}%
\end{figure}

The saturation curves  for porcolation simulations on Sierpi\'nski triangle with levels 8-13 are presented in Figure \ref{fig:sierpenski-triangle-sat}. The percolation thresholds are significantly higher than for simple square networks.
This can be explained by a special characteristic of the Sierpi\'nski triangle. 
Some vertices (highlighted in Figure \ref{fig:sierpenski-triangle-4} with lighter color and bigger size) are critical from the perspective of porcolation, since there is no other path to reach the new region. These vertices form an articulation set, since if they are removed, the graph falls apart, hence it is not a robust graph.

\begin{figure}[h]
	\centering
	\subfigure[]{
		\includegraphics[width=0.28\textwidth]{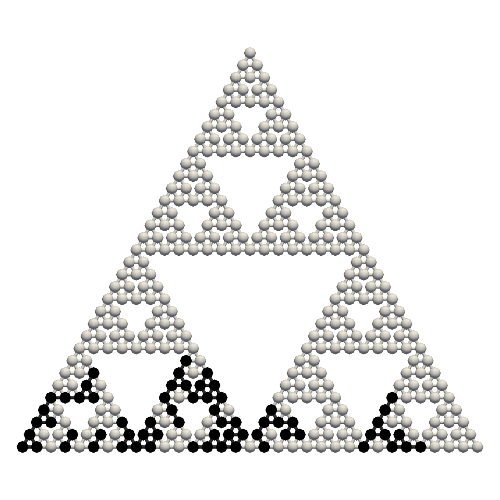}
		\label{fig:sierpenski-triangle-p3}%
	}
	~
	\subfigure[]{
		\includegraphics[width=0.28\textwidth]{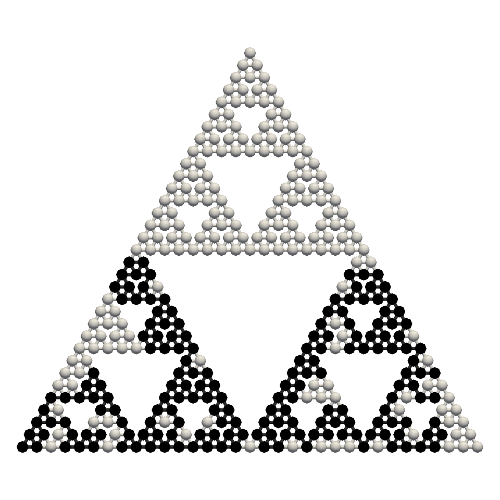}
		\label{fig:sierpenski-triangle-p6}%
	}
	~
	\subfigure[]{
		\includegraphics[width=0.28\textwidth]{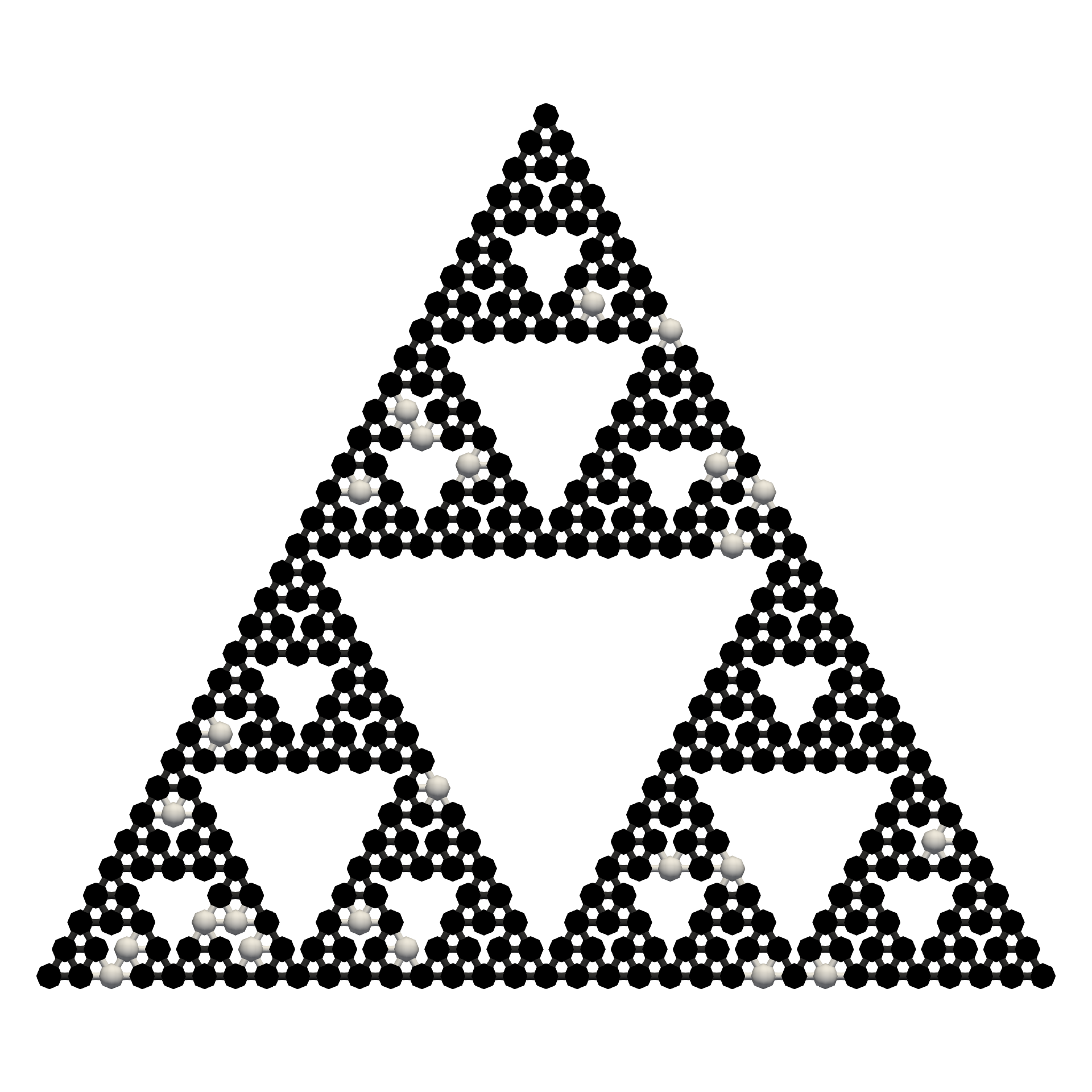}
		\label{fig:sierpenski-triangle-p8}
	}
	\caption{Fluid distribution of Sierpi\'nski triangle with level 5 for different time steps, the corresponding dimensionless pressures are:  (a) $p=0.4$, (b) $p=0.7$, and (c) $p=0.9$. The gray vertices are empty, the black vertices are the occupied ones. }%
	\label{fig:sierpenski-triangle-time}%
\end{figure}

We also observe that the saturation curves are shifted towards the $p=1$ dimensionless pressure as the graph level is increased. 
The reason for this shift is also connected to the articulation set, since as the graph level is increased the number of such critical vertices is also increasing. The inflection points (corresponding to the percolation thresholds $P_\mathrm{crit}$) of the saturation curves are shown in Figure \ref{fig:sierpenski-triangle-inf}.
The occupation of sites for the Sierpi\'nski triangle graph are shown in Figure \ref{fig:sierpenski-triangle-time} at different time steps.

\begin{figure}[h]
	\centering
	\subfigure[]{
		\includegraphics[width=0.6\textwidth]{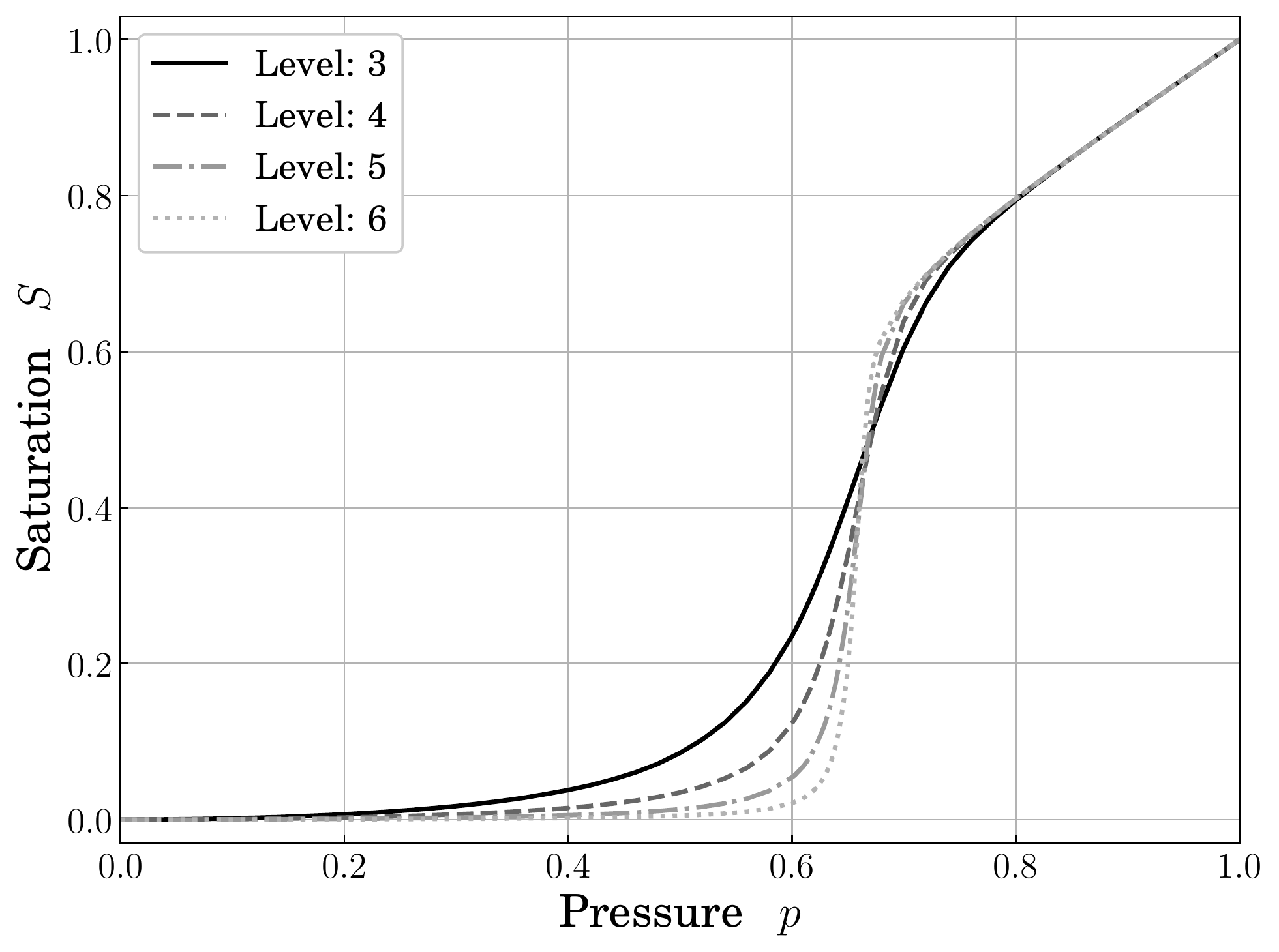}
		\label{fig:sierpenski-carpet-sat}%
	}
	~
	\subfigure[]{
		\includegraphics[width=0.3\textwidth]{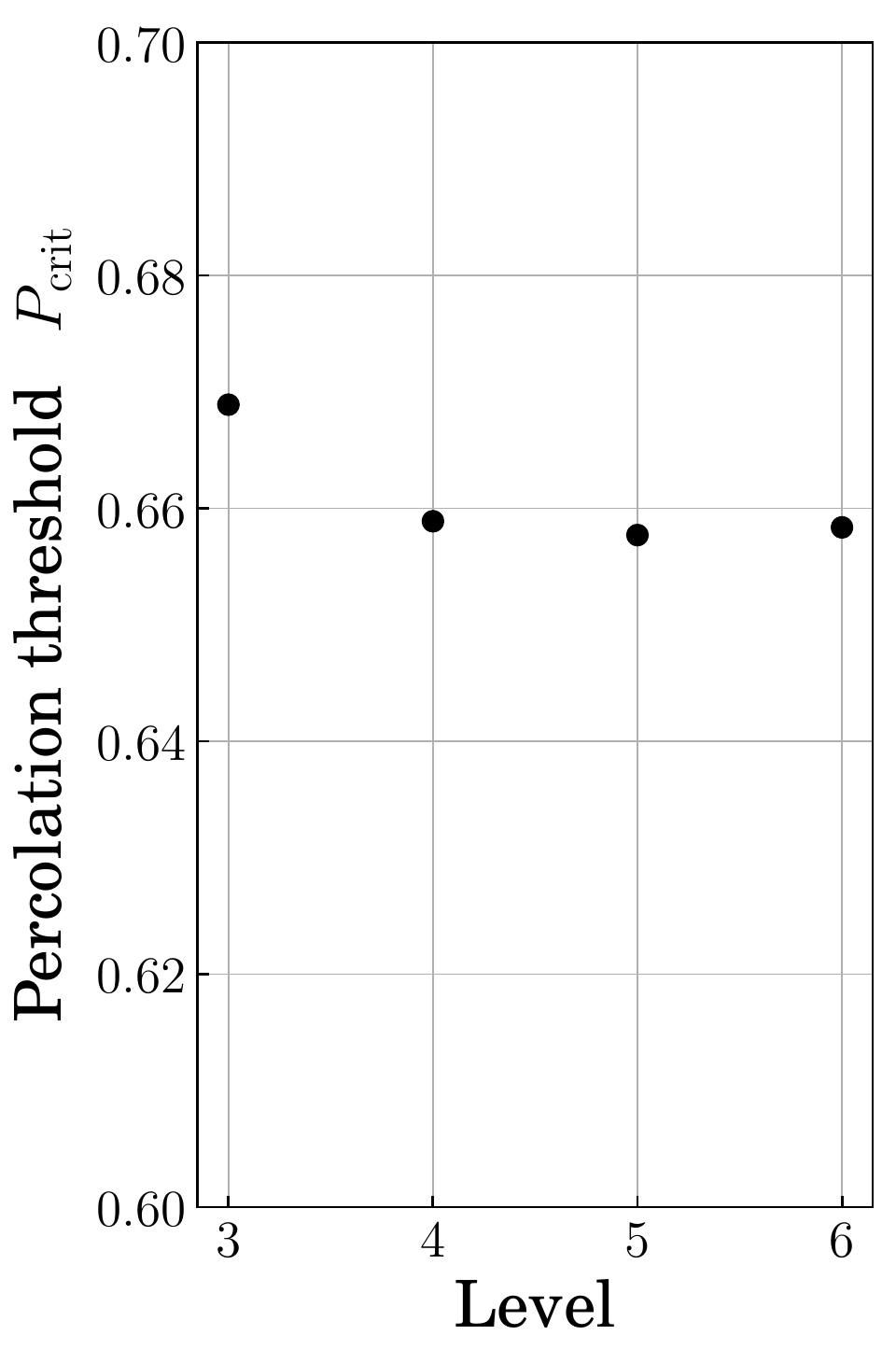}
		\label{fig:sierpenski-carpet--inf}
	}
	\caption{(a) Saturation curves for Sierpi\'nski carpets of different levels, (b) Percolation thresholds vs. graph level }%
	\label{fig:sierpenski-carpet-results}%
\end{figure}

\begin{figure}[h]
	\centering
	\subfigure[]{
		\includegraphics[width=0.27\textwidth]{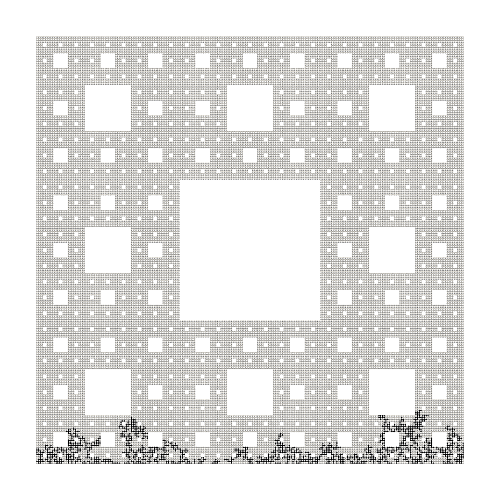}
		\label{fig:sierpenski-carpet-p3}%
	}
	~
	\subfigure[]{
		\includegraphics[width=0.27\textwidth]{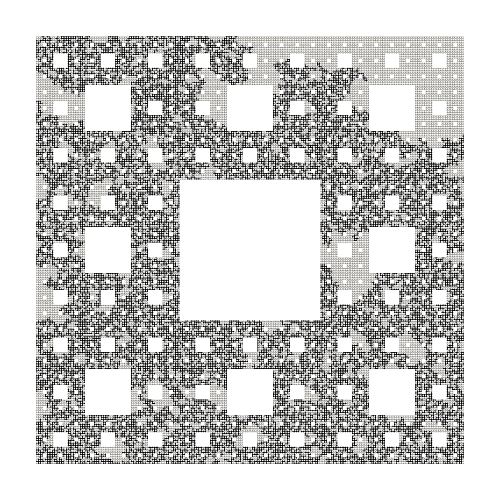}
		\label{fig:sierpenski-carpet-p5}%
	}
	~
	\subfigure[]{
		\includegraphics[width=0.27\textwidth]{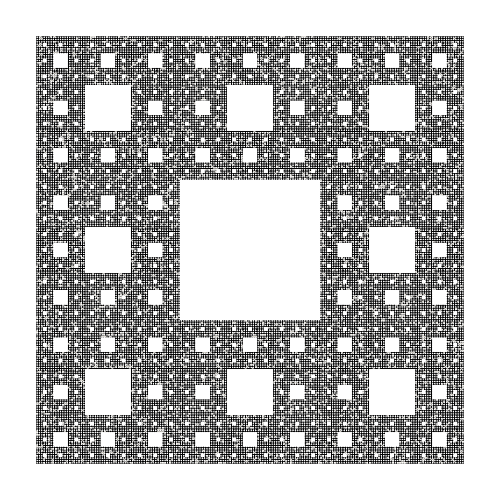}
		\label{fig:sierpenski-carpet-p7}
	}
	\caption{Fluid distribution of Sierpi\'nski carpet with level 5 for different time steps, the corresponding dimensionless pressures are:  (a) $p=0.5$,  (b) $p=0.6$,  and (c) $p=0.9$. The gray vertices are empty, the black vertices are the occupied ones.}%
	\label{fig:sierpenski-carpet-time}%
\end{figure}

Saturation curves of Sierpi\'nski carpet with levels 3-6 are shown in Figure \ref{fig:sierpenski-carpet-sat}. 
These curves are not shifted towards $p=1$ dimensionless pressure as the graph level is increased because these graphs are more robust.
The percolation thresholds remain almost the same for different graph levels, they are all in the range [0.65,~0.67] as shown in \ref{fig:sierpenski-carpet--inf}. 
The occupation of sites for the Sierpi\'nski carpet graph are shown in Figure \ref{fig:sierpenski-carpet-time} at different time steps.




\clearpage
\section{Porcolation on random graphs}\label{section:irregular-graphs}

Real porous medium has irregularly distributed  pores, therefore we also studied porcolation on irregular (random) networks with arbitrary pore degree
distributions.

There are two fundamental ways to generate random networks: edge cutting (removing) and edge adding methods. 
The edge cutting method starts from an existing network and removes edges until the prescribed pore degree distribution is reached.
The edge adding method starts from a set of nodes and adds edges until the prescribed pore degree distribution is obtained. Due to its flexibility the edge adding method was chosen.

\subsection{Random pore network generation}

The so-called ``configuration model'' \cite{britton2006generating} creates random networks of a given pore degree distribution.
First, the desired number of pores are created with assigned coordination number (pore degree) with the given distribution. 
The prescribed pore degree at each pore can be imagined as attached ``stubs''. Randomly chosen stubs are connected until each vertex has the prescribed number of neighbors.
In real porous media usually only spatially close  pores are connected, but this characteristic is not taken into account in the configuration model. 
A pure graph model does not contain an inherent distance metric, a Euclidean graph (a pure graph embedded into Euclidean space) is the appropriate object to represent a real pore-throat network.

\begin{figure}[h]
	\centering
	\includegraphics[width=0.7172\textwidth]{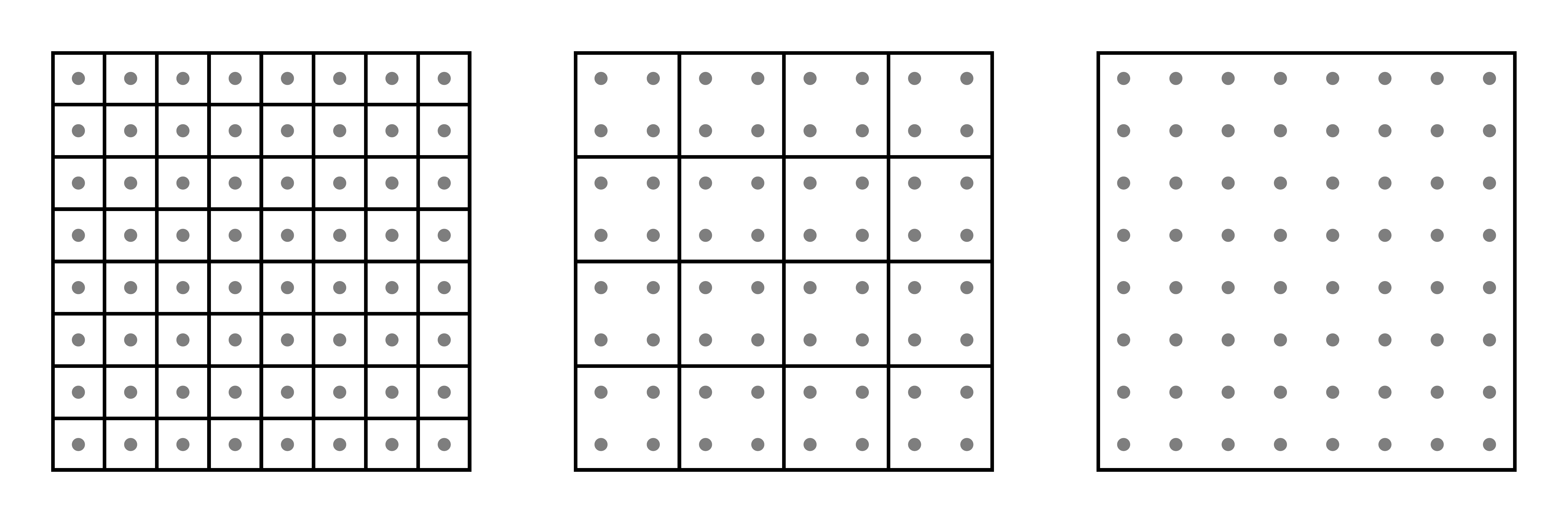}
	\caption{Possible cases with the modified cell list algorithm: one pore in each cell (left), some pores in each sell (middle), and all pores in one cell (right)}%
	\label{fig:cell_list_extreme}%
\end{figure}

To efficiently generate Euclidean pore networks we developed a modified version of the cell list algorithm \cite{allen1990computer}. 
In this method the 3-dimensional Euclidean space is partitioned into non-overlapping cells (for computational simplicity our cells were cubes). Two cells are called neighbors if their intersection has a positive area. 
A given number of points (representing pores) are added to each cell. 
Throats are then added to  connect pores only in neighboring cells. 

There are two extreme cases of the modified cell list algorithm: when only one cell is defined, and when every cell contains only one pore. 
The single cell case is equivalent to the configuration model, while the case when every pore has it's own cell is equivalent to the simple cubic network. 
These extreme cases are depicted in 2D on the right and left side of Figure \ref{fig:cell_list_extreme}. 
The middle part shows the 4 pores/cell setup of modified cell list algorithm.
This method is capable of generating 3D graphs as well.

The pure graph created by the configuration model is a ``global'' network because there is no spatial restriction for the neighboring pores, while the one made with the developed cell list algorithm is a ``local''  network because only the spatially close pores can be connected.

\subsection{Influence of locality on saturation}

We can imagine a porous rock sample as pores and throats connecting them.
Connected pores generally are not too far from each other, this is the localized nature of real pore networks.
Graph locality can be quantified by the statistics of pores in each cell. 
We examined the effect of graph locality in porcolation simulations using the introduced modified cell list algorithm for graph generation. 

\begin{figure}[t]
	\centering
	\includegraphics[width=0.7\textwidth]{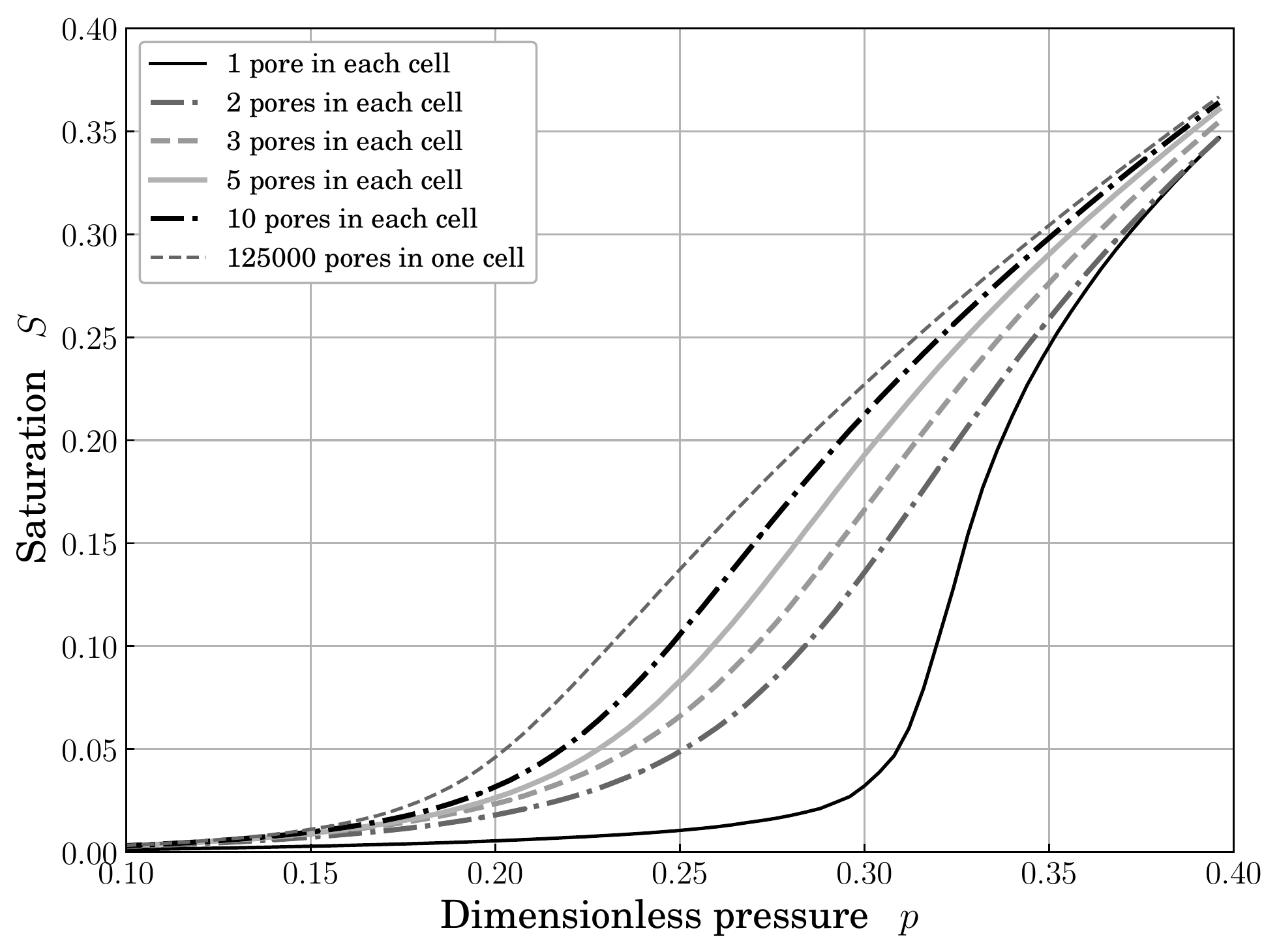}
	\caption{Saturation curves for examining the effect of graph locality.}%
	\label{fig:local_sat_curves}%
\end{figure}

Simulations were performed on $50^3$ cubic networks (125000 pores) with one invading side and the results were averaged over 100 simulations. 
The entry pressure values were generated from uniform distribution in the range of [0,\,1].
The 1, 2, 3, 5, 10, and 125000 pores/cell setups were investigated, these resulted in 6, 13, 20, 34, 69, and 124999 possible pore neighbors, respectively.
The saturation curves are shown in Figure~\ref{fig:local_sat_curves}.

As we see, network locality has a significant role in porcolation, more local pore
network means higher percolation threshold values. 
We also conclude that the modified cell list
algorithm is capable of generating more realistic pore networks than the original Britton \textit{et al.} algorithm. 

\subsection{Experiments with different pore degree distributions}

\begin{figure}[h]
	\centering
	\includegraphics[width=0.7\textwidth]{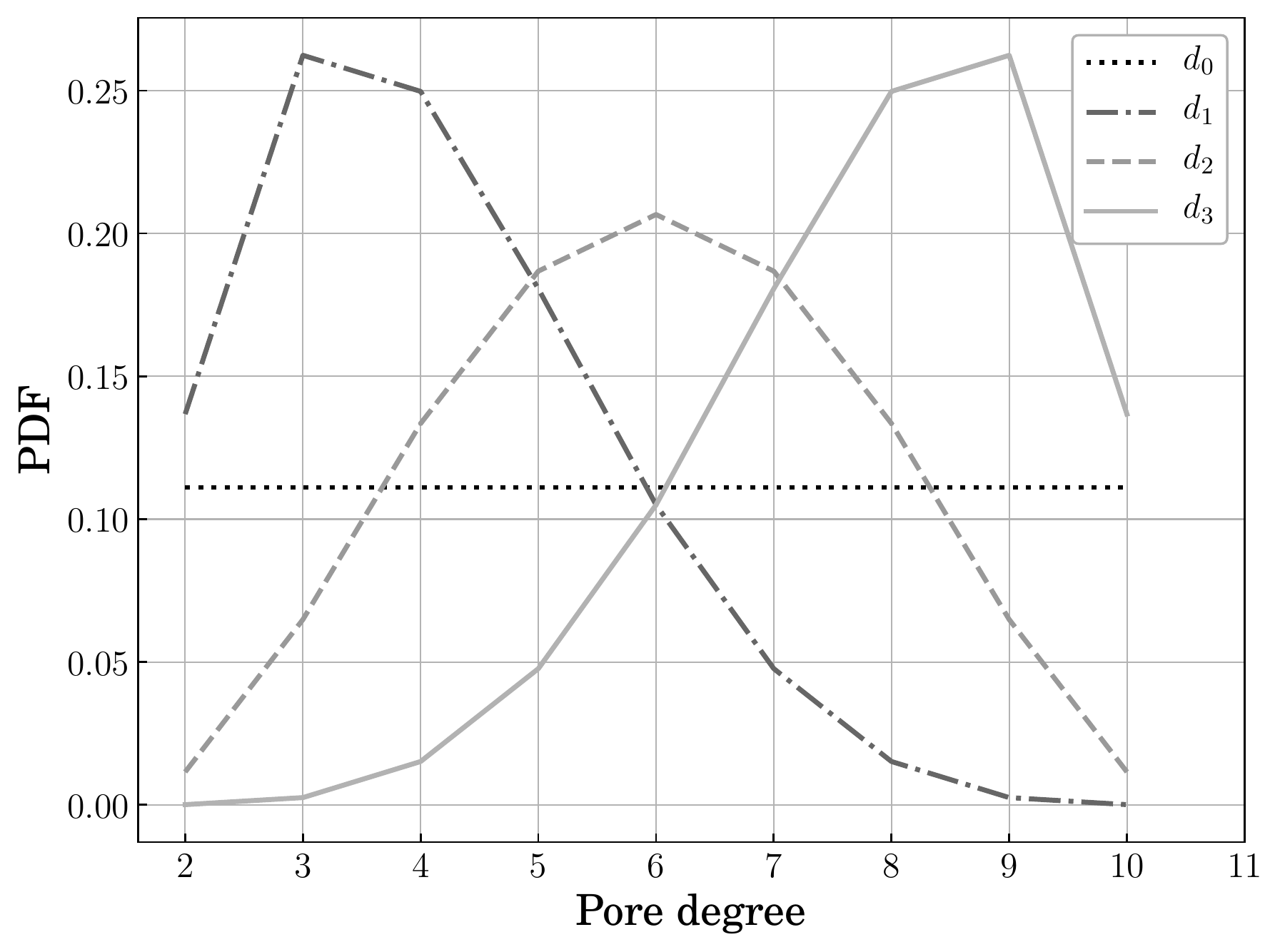}
	\caption{The discrete pore degree distributions}%
	\label{fig:pore_degree_dist}%
\end{figure}

The modified cell list algorithm is also capable of generating random networks with a prescribed pore degree distribution. 
It was tested with 4 different pore degree distributions: $d_0$ is the uniform distribution, $d_1$ emphasizes low pore degrees, $d_2$ embraces the middle range and $d_3$ is a distribution where the high pore degrees are dominant (see Figure \ref{fig:pore_degree_dist}).

\begin{figure}[h]
	\centering
	\includegraphics[width=0.7\textwidth]{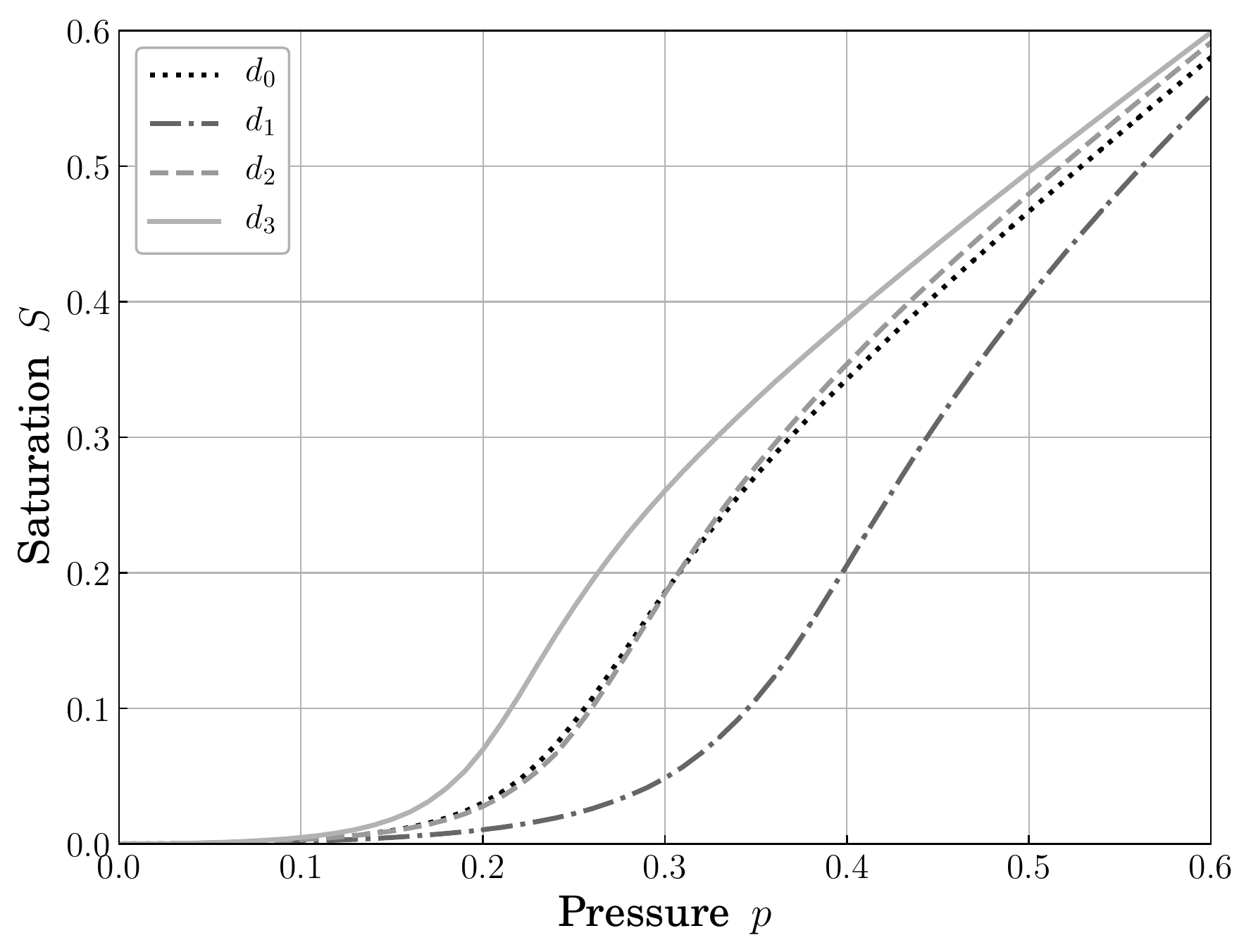}
	\caption{Saturation for different pore degree distributions}%
	\label{fig:pore_degree_sat}%
\end{figure}

The 4 pore/cell setup was used for the modified cell list algorithm to generate the pore networks.
Simulations were performed on $50^3$ networks with one invading side and the results were averaged over 100 simulations, as before. 
The entry pressure values were generated from uniform distribution in the range of [0,\,1].
The saturation curves are shown in Figure~\ref{fig:pore_degree_sat}.

The results for the $d_0$ (uniform) and the $d_2$ (middle dominant) pore degree distributions are almost the same.
The results for the $d_1$ (low dominant) and the $d_3$ (high dominant) pore degree distributions are remarkably different as we expected. This is caused by the high difference between the number total connection in the pore networks.

\clearpage
\section{Conclusion}\label{section:conclusion}

In order to obtain saturation curves for different networks we implemented the porcolation model in OpenPNM with the built-in drainage simulation.
The first porcolation simulations were carried out on square and cubic networks to validate the model: the inflection points of the saturation curves correspond well with the theoretical percolation threshold values.

The saturation curves were also determined for networks based on Sierpi\'nski triangle and Sierpi\'nski carpet. 
If the graph level of the Sierpi\'nski triangle is increased, the inflection of the saturation curve is shifted to the higher dimensionless pressure values. 
This phenomenon is caused by the increasing number of vertices in the articulation set, which removal would disconnect the graph.

We developed a network generation method (based on the cell list algorithm) which is capable of efficiently generating local pore networks with random pore degree distribution. 
We showed that the locality of pore networks has major effect on the saturation curves and we also done porcolation simulations on pore networks with random pore degree distributions.

\bibliographystyle{unsrt}
\bibliography{saturationpaper}

\end{document}